\documentclass[lettersize,journal]{IEEEtran}

\usepackage{xspace}
\usepackage{enumitem}
\usepackage[normalem]{ulem}
\usepackage{tabularx} 
\usepackage{amsmath}
\usepackage{amssymb}
\usepackage[caption=false,font=normalsize,labelfont=sf,textfont=sf]{subfig}
\usepackage{orcidlink}
\usepackage{xcolor}
\usepackage{soul}
\usepackage{color}
\usepackage{array}
\usepackage{colortbl}
\usepackage{longtable}
\usepackage{enumitem} 
\usepackage{orcidlink}
\useunder{\uline}{\ul}{}
\usepackage{pifont}% http://ctan.org/pkg/pifont
\usepackage{multirow}
\usepackage{makecell}
\newcommand{\qq}[1]{``#1''}
\usepackage{makecell}
\usepackage{comment}
\usepackage{array}
\usepackage[table,xcdraw]{xcolor}
\usepackage{balance}
\usepackage{colortbl}
\usepackage{rotating}
\usepackage{float}
\definecolor{happy}{HTML}{2ECC71}   % Bright green
\definecolor{anger}{HTML}{E74C3C}  % Red
\definecolor{fear}{HTML}{8E44AD}   % Purple
\definecolor{disgust}{HTML}{B7950B}% Brown
\definecolor{sad}{HTML}{3498DB}    % Blue
\definecolor{neutral}{HTML}{95A5A6} % Gray

\newcommand{\system}[1]{\emph{MAJIC}\xspace}

\author{
Tanmay Srivastava\,\orcidlink{0000-0003-0144-7931},
Paras Bhavnani\,\orcidlink{0009-0004-8656-3257},
Benjir Alvee Islam\,\orcidlink{0009-0002-3624-4923},
and Shubham Jain\,\orcidlink{0000-0002-4864-6420}%
\thanks{
The authors are with Stony Brook University,
Stony Brook, NY, USA.
}
}
\begin{document}

% \title{Enhancing Speech-based Emotion Recognition using Articulatory Motion}
% \title{Beyond Acoustics: Multimodal Emotion Recognition Through Articulatory Motion Sensing} ArtEmo
% \title{Speech Meets Motion: Advancing Emotion Recognition with Articulatory Motion}
\title{\emph{MAJIC}: Leveraging Articulatory Motion for Speech-based Emotion Recognition}

\maketitle

\begin{abstract}
\textcolor{black}{We introduce ~\system{}, a multimodal emotion recognition system that leverages articulatory motion of the jaw and facial muscles for speech-based emotion recognition (SER). While most SER systems perform well on datasets with strongly expressed emotional speech of trained actors, their performance often degrades when emotional expressions become more subtle. We explore this challenge by engineering features from articulatory motion and integrating them with audio features using a multi-task learning framework.
Our key insight is that emotion in speech manifests not only through vocal characteristics but also through distinct articulatory motions: jaw movements, facial muscle vibrations, and speech-induced vibrations. While audio captures features such as pitch and prosody, articulatory motion contains complementary information that is not present in audio alone. 
 We evaluate our system on data collected from 20 participants across multiple sessions, 10 languages, and diverse scenarios, including prompted and conversational speech, showing its robustness across users and settings.~\system{} achieves 93\% accuracy and 91\% F1 score for emotion classification, outperforming strong audio-based baselines on our dataset.}

\end{abstract}

%\begin{IEEEkeywords}
%Emotion recognition, articulatory motion sensing, multi-task learning, wearable computing, and affective computing.
%\end{IEEEkeywords}

\section{Introduction}

Human emotion recognition has emerged as the next big challenge for Artificial Intelligence (AI) agents, particularly virtual assistants and conversational agents that increasingly rely on speech as their primary interaction modality. 
In human-computer interaction (HCI) and robotics, emotion recognition leverages acoustic features to enable adaptive virtual assistants and interactive environments that respond intelligently to the speaker's emotional states~\cite{alnsour2026ai}. For healthcare applications, these systems support mental health monitoring by detecting emotional patterns through conversational speech in real-time, facilitating continuous passive assessment for conditions like depression or anxiety. In customer service contexts, emotion recognition capabilities enhance AI systems by enabling more empathetic, contextually appropriate interactions with users, as they analyze vocal signals to detect user frustration. 

This has fueled advances in speech emotion recognition (SER) techniques 
~\cite{ma2024emotion2vec, ben2024enhancing, tyagi2024exploring}, achieving high accuracy on benchmark datasets like IEMOCAP~\cite{busso2008iemocap} and RAVDESS~\cite{livingstone2018ryerson}. 
However, a fundamental limitation persists: these systems are predominantly trained and evaluated on data collected from professional actors who deliver emotionally expressive speech that is clear, exaggerated, and consistent, qualities gained through years of training. As a result, current SER models often show reduced performance on emotional expressions of non-actors that have subtle and less exaggerated features.
Our preliminary analysis reveals that state-of-the-art audio-based and audio-text SER systems have an accuracy drop of over 30\% when transitioning from actor-based speech to speech from non-actor users (Table~\ref{tab:sota_res}). This gap highlights a key challenge in deploying emotion recognition for next-generation AI systems. For SER to be broadly applicable, systems must be able to interpret a wider range of emotional expressions, including those that are less exaggerated than actor-performed speech.

Recognizing this limitation, researchers have explored multi-modal approaches that combine audio with complementary modalities to improve emotion recognition. Text-based methods extract semantic features from speech transcripts to complement acoustic analysis, but rely heavily on linguistic content and often miss critical emotional nuances conveyed through prosody~\cite{acheampong2020text, alswaidan2020survey}. Vision-based systems integrate facial features with audio, but suffer from occlusion issues, lighting variations, and privacy concerns in real-world uses~\cite{Kjeldsen52:online, abate2023limitations}. Other efforts have incorporated physiological signals such as EMG, ECG, body posture, and respiratory patterns to capture broader physiological manifestations of emotional states~\cite{mittal2020m3er, koelstra2011deap, soleymani2011multimodal, abadi2015decaf}, yet these require extensive sensor arrays, specialized equipment, and complex calibration, limiting scalability. 
Electromagnetic articulography (EMA) systems have shown that articulatory motion encodes emotional information -- jaw opening degree increases with annoyance, lateral lip distance correlates with emotional expression, and articulation-based features can outperform acoustic features for emotion classification~\cite{erickson1998articulatory, nordstrand2004measurements}. 
However, these works require high-fidelity approaches that demand expensive laboratory infrastructure~\cite{shah2019articulation,lee2006study}. 
The fundamental challenge remains: \emph{current multi-modal solutions improve accuracy but rely on sensing modalities that are expensive, privacy-sensitive, or difficult to deploy in everyday settings.}

To address these limitations, we present~\system{}, a \textit{M}ulti-modal \textit{A}udio and \textit{J}aw \textit{I}ntegration approach for emotion \textit{C}lassification. Our key insight is that jaw motion provides complementary information for SER beyond acoustic features, and can be captured using lightweight, non-invasive wearable IMU sensors.
~\system{} employs a single IMU sensor placed near the temporomandibular joint (TMJ) to track jaw motion patterns that encode aspects of speech production. Unlike previous multi-modal solutions, our approach does not require specialized laboratory equipment, invasive sensor placement, or privacy-sensitive modalities. To the best of our knowledge,\emph{~\system{} is the first wearable IMU-based approach that leverages articulatory motion sensing near the TMJ for emotion recognition.} We capture both pre-speech articulatory dynamics, which reveal emotion-related cues before vocalization, and speech-related articulatory features, providing information that is unavailable to audio-only systems.

\textbf{Challenges}: There are three challenges in learning emotional cues from audio and jaw motion for non-actors: (1) The jaw is a secondary articulator and there is limited literature exploring its role in emotional expression. Developing methods to accurately map jaw motion to emotional states while minimizing noise or artifacts is therefore challenging. (2) State-of-the-art audio-based emotion recognition systems often rely on datasets collected from professional actors, who produce exaggerated emotional expressions. In contrast, we collect data from non-actor participants expressing emotions, where variations in acoustic features across emotions are more subtle. (3) Emotions are inherently continuous and exist on a spectrum, making it difficult to discretize them into well-defined categories with clear boundaries. This complicates the encoding of emotions and makes recognition challenging.

\textbf{Contributions}: In this paper, we make the following contributions:
\begin{itemize}[leftmargin=*,topsep=0pt]
\item We develop~\system{}, a wearable IMU-based system that integrates articulatory motion sensing with audio for speech emotion recognition. Our approach is language and content-agnostic, enabling use across multilingual settings.
\item We analyze pre-speech jaw motion patterns, capturing preparatory articulatory behavior that occurs before speech begins and is not present in audio-only approaches.
\item We extract articulatory features based on motion characteristics such as sharpness, smoothness, and temporal patterns that are associated with different emotional expressions.
\item We evaluate~\system{} with 20 users across multiple scenarios, including different phrases, same phrases, multiple languages, and conversational responses. Our system outperforms strong audio-based baselines on our dataset while requiring minimal training data from each user.
\end{itemize}

\system{} is a step toward practical multimodal SER systems by combining articulatory motion with acoustic features.

% \textcolor{blue}{Changes made:}

% \begin{itemize}
%     \item Removed real-world emotion, natural emotion, and strong attack on prior works
%     \item Softened claims “fail”, “reduced performance”, “significant advancement”  “step toward”
%     \item Fixed dataset narrative. now: “non-actor participants expressing everyday emotional speech”
%     \item Fixed novelty claim “first system”  “first wearable IMU-based approach…”
% \end{itemize}
\begin{table}[t]
\centering
\begin{tabular}{|l|c|c|}
\hline
 & \textbf{Actors} & \textbf{Non-Actors}   \\ \hline
SOTA SER~\cite{ma2024emotion2vec} & 85\%                 & 65\%              \\  \hline
Audio+Text~\cite{radford2022robust, savani_bert_emotion}  & 88\%                & 44.3\%            \\  \hline
\end{tabular}
\caption{Performance comparison of SER with professional actors versus non-actors.}
\vspace{-0.5cm}
\label{tab:sota_res}
\end{table}

\section{Motivation and Opportunities}
\label{sec:bg}
Emotion is a significant psychological state in human behavior.  A dimensional model, referred to as the Circumplex Model~\cite{russell1980circumplex} is used to conceptualize emotions in a two-dimensional space categorized by valence and arousal. Valence refers to the emotional state of mind, such as positive (happy) or negative emotions (sad), and arousal is the intensity of emotion: high activation (angry) and low activation (sad).

\begin{figure}
    \centering
    \includegraphics[width=\linewidth]{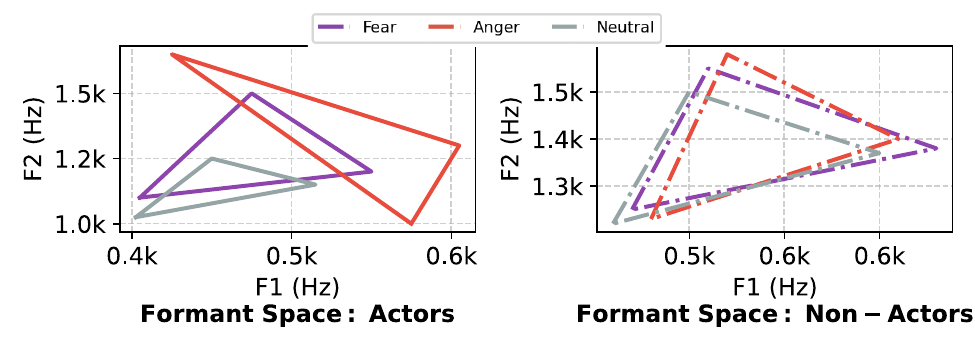}
    \vspace{-0.5cm}
    \caption{Formant triangles for actors show distinct emotional patterns (F1-F2 space), while for non-actors, it reveals significant overlap between emotions.}
    \label{fig:formants}
    
\end{figure}

\begin{figure}
    \centering
    \includegraphics[width=\linewidth]{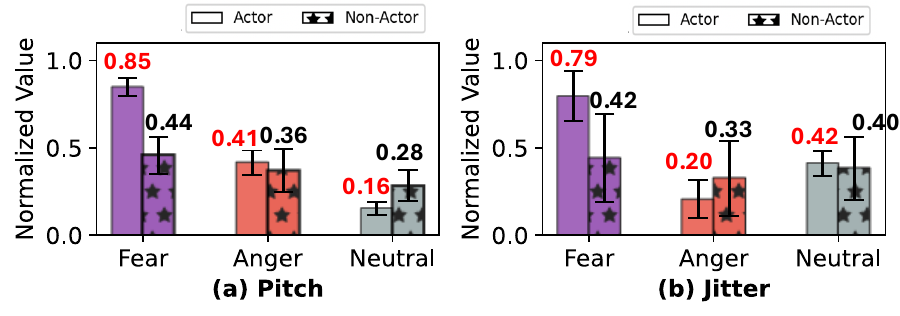}
    \vspace{-0.5cm}
    \caption{Normalized acoustic features. Both Pitch and Jitter exhibit a much larger range for actors. }
    \label{fig:pitchNjitter}
    \vspace{-0.5cm}
\end{figure}

\subsection{Gap in current SER methods}
\label{sec:motivation}

Our primary motivation for this work stems from the observation that existing SER works 
\textcolor{black}{often show reduced performance} 
for users with no professional training (\emph{non-actors}). 
To investigate this gap, we collected data from non-actors across various emotional states (\S\ref{sec:data}) and evaluated a state-of-the-art SER approach~\cite{ma2024emotion2vec}. As shown in Table~\ref{tab:sota_res}, this approach~\cite{ma2024emotion2vec} achieves approximately 85\% accuracy when classifying emotions expressed by professional actors but drops to 65\% for non-actors.
Similarly, combining audio recognition~\cite{radford2022robust} with BERT-based emotion analysis~\cite{savani_bert_emotion} exhibits an even larger disparity, dropping from 88\% accuracy with actors to just 44\% with non-actors. 
This substantial accuracy reduction reflects a fundamental challenge: 
\textcolor{black}{less exaggerated emotional expressions often exhibit fewer consistent patterns compared to acted performances,} 
exhibiting more subtle cues instead.
Figure~\ref{fig:formants} shows how \emph{formants}, a widely used audio-based feature for SER, differs for actors and non-actors. We can see that the formant space for non-actors shows significant overlap across emotions in contrast to the distinguishable distribution for actors. Similarly, for audio-based features, such as pitch and jitter shown in Figure~\ref{fig:pitchNjitter}, the range for actors is much larger than for non-actors across emotions. 
There is a need for augmenting audio data with other modalities to capture the subtle variations in emotions exhibited by non-actors. Inspired by recent work~\cite{srivastava2022muteit,srivastava2023jawthenticate,srivastava2024unvoiced,shi2021face} in leveraging human articulatory motion, we investigate their potential importance in enhancing emotion recognition
\footnote{It is to note that these observations are based on prompted expressions and are intended to highlight differences in expression characteristics rather than fully natural emotional behavior.}.

\subsection{Role of articulators in emotional speech}

Emotional states influence speech production, affecting acoustic characteristics such as pitch, intensity, speech rate, and rhythm~\cite{thoughtco_prosody,ooi2014new,ozseven2018acoustic}. Since these acoustic properties arise from articulatory behavior, emotional expression is also reflected in the motion of speech articulators.

\noindent \textbf{Articulatory motion.}
Prior work has shown that articulatory motion encodes emotion-related information~\cite{shah2019articulation,thoughtco_prosody,busso2004analysis}. In particular, emotional states are associated with distinct jaw motion patterns and facial muscle activity~\cite{Ostry1999CONTROLOC,lee2005articulatory}. For example, happy speech often exhibits wider jaw motion, while fear and anger are associated with characteristic motion variations such as tremors or abrupt movements~\cite{asai2017articulatory,ren2018articulatory}. These observations suggest that articulatory motion provides complementary cues for emotion recognition.

\noindent \textbf{High-frequency IMU components and facial vibrations.}
Speech production generates facial vibrations through articulatory muscle activity and speech-induced mechanical vibrations transmitted through facial tissues and bone~\cite{srivastava2023jawthenticate}. Although these signals do not directly encode acoustic features such as pitch or formants, they capture biomechanical aspects of speech production. Because emotions influence muscle tension and vocal effort, the resulting vibration patterns vary across emotional states, particularly for high-arousal emotions such as happiness and anger~\cite{ali2018emotion}. These observations suggest that facial vibrations and high-frequency IMU components provide information complementary to conventional acoustic features.

\begin{figure}
    \centering
    \includegraphics[width=\linewidth]{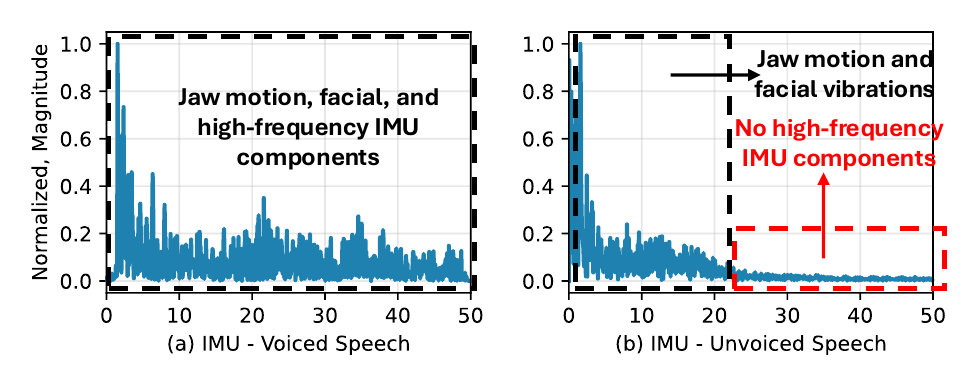}
    \vspace{-0.5cm}
    \caption{Frequency spectra of articulatory motion captured by IMU and EMG sensors during speech. During voiced speech (a), IMU captures jaw motion (0-5Hz), facial (5- 20Hz), and high-frequency IMU components (20- 50Hz). During unvoiced speech (b), high-frequency IMU components (20- 50Hz) are largely absent while jaw and facial components persist.}
    \label{fig:diff_freq_bands_emg}
\end{figure}

{}

\subsection{Opportunities and Preliminary Study}
We see advancements in articulatory motion analysis as an opportunity to make SER accessible to everyday users (non-actors) by augmenting speech with semantic features from jaw motion and facial vibrations. As a first step, we identify the specific frequency bands where key articulatory components (jaw motion, high-frequency IMU components, and facial muscle activity) are most prominent. Using an example where a participant articulated the phrase \textit{\qq{Have a good day}} in both voiced (with vocalization) and unvoiced (without vocalization) manners, shown in Figure~\ref{fig:diff_freq_bands_emg}, we observe the presence of additional high-frequency ($>$20 Hz) components in Figure~\ref{fig:diff_freq_bands_emg}(a) that are absent in Figure~\ref{fig:diff_freq_bands_emg}(b). These higher-frequency components observed during voiced articulation arise from speech-induced mechanical vibrations that propagate through facial tissues and bone~\cite{shi2021face}. Prior works have shown that gross jaw motion is primarily concentrated below 5 Hz, while facial muscle activity and associated facial vibrations are typically observed in the 5--20 Hz range~\cite{srivastava2022muteit,jawsense}. 

\begin{figure}
    \centering
    \includegraphics[width=1.05\linewidth]{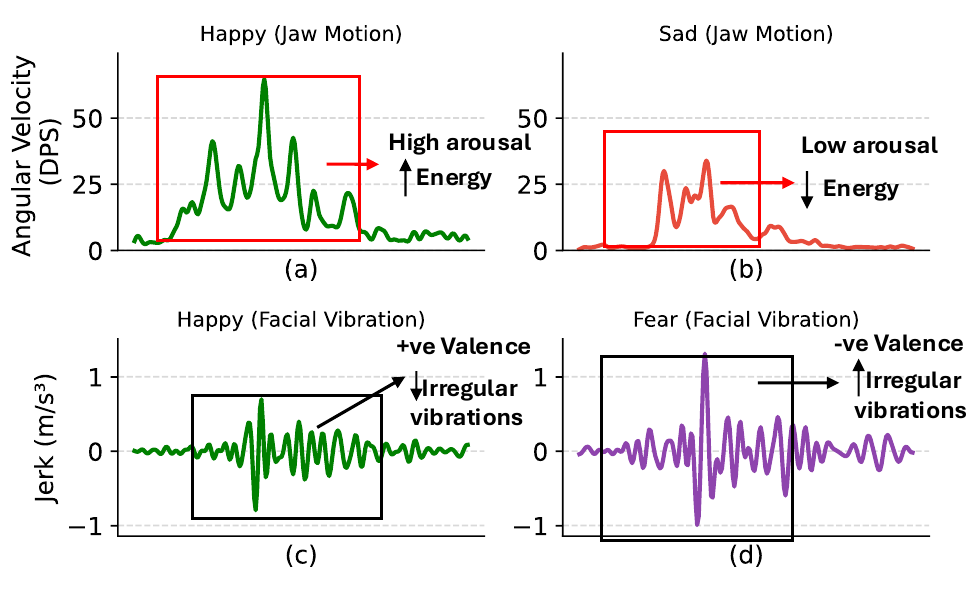}
    \vspace{-0.5cm}
    \caption{\emph{\qq{I cannot believe this.}} in different emotions. Jaw motion shows (a) higher energy for \emph{Happy} than (b) \emph{Sad}. Jerk in facial vibrations (5-20Hz) shows (c) \emph{Happy} with smoother patterns versus (d) \emph{Fear} with more abrupt acceleration changes.}
    \vspace{-0.5cm}
    \label{fig:poc}
\end{figure}

Having isolated the signals, we analyze how emotional states affect each articulatory component.
In jaw motion (0–5 Hz), happy speech exhibits higher angular velocity than sad speech (Figure~\ref{fig:poc}(a,b)), consistent with the higher arousal of positive emotions. In facial vibrations (5–20 Hz), happy jerk patterns (Figure~\ref{fig:poc}(c)) are smoother and more regular, indicating positive valence, while fear (Figure~\ref{fig:poc}(d)) shows abrupt, irregular patterns despite also being high arousal. These observations illustrate how arousal and valence manifest across different articulatory frequency ranges.
Since higher-frequency IMU components share similar acoustic properties with audio signals, rather than oversampling the IMU for acoustic features as in prior work~\cite{10.1145/3699757}, we focus on lower-frequency bands (0–50 Hz) for emotion analysis and capture high-quality audio separately via a microphone.

These observations establish that emotional states manifest in both modalities—audio and articulatory motion. We investigate whether lightweight sensing can capture complementary signals that improve emotion recognition. These cues are obtainable from a non-invasive IMU near the TMJ, offering a privacy-preserving alternative to cameras and a less intrusive solution than EMG. This motivates our key question: \emph{\qq{How can we effectively combine articulatory and audio features to achieve robust emotion classification for everyday users?}}

% \vspace{-0.4cm}
\section{System Overview}

\begin{figure}
    \centering
    \includegraphics[width=0.8\linewidth]{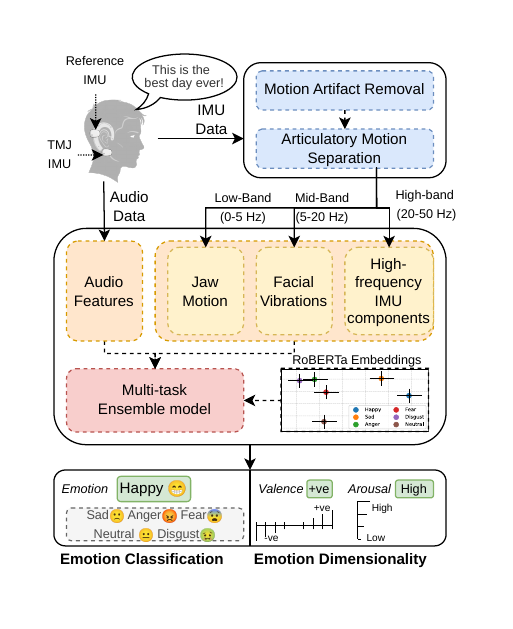}
    \vspace{-1cm}
    %\vspace{-0.5cm}
    \caption{Overview of \system{}. We extract articulatory and audio features, combine them with semantic relationships in a multi-task model for recognition.}
    \label{fig:system_overview}
    % \vspace{-0.5cm}
    \vspace{-0.5cm}
\end{figure}

We develop a multimodal framework that leverages both the jaw motion patterns across different frequency bands and the corresponding acoustic properties to recognize emotions from speech.
Figure~\ref{fig:system_overview} shows the system overview with three main processing steps.
First, we preprocess the raw IMU signal to isolate distinct vibration patterns that occur during speech. The IMU captures three types of motion: slow articulatory jaw motions that reflect speech patterns (0-5 Hz), mid-frequency facial muscle vibrations that indicate skin deformation patterns (5-20 Hz), and high-frequency (20-50 Hz) 
\textcolor{black}{speech-induced vibrations associated with higher-frequency articulatory activity}. 
We use the twin-IMU setup from previous works~\cite{srivastava2023jawthenticate, srivastava2024unvoiced} to mitigate the effect of body movements that could pollute jaw motion.

Second, we extract representative features from the processed IMU and audio signals. From jaw motion, we capture temporal patterns like sharpness and activity bursts. Facial vibrations provide frequency-domain features that 
\textcolor{black}{capture variations associated with different emotional expressions}. 
\textcolor{black}{Higher-frequency IMU components provide complementary motion information that is not present in audio alone.}
The audio processing extracts prosodic and spectral features that capture variations in speech, such as pitch and energy patterns characteristic of different emotions.

Finally, we design a multitask learning model that combines these multimodal features with text (RoBERTa) embeddings to classify emotions. The text embeddings help the model 
\textcolor{black}{capture semantic relationships between emotion labels}, 
for instance, fear and disgust are more closely related than happiness and sadness. 
\textcolor{black}{This provides additional contextual information during training (Section~\ref{sec:recog}).}
The model simultaneously outputs the emotion category (happy, sad, angry, disgust, fear, neutral), the valence (+/-), and the intensity of the arousal ($\uparrow$/$\downarrow$).

\begin{figure*}[hbt!]
    \centering
    \includegraphics[width=0.9\linewidth]{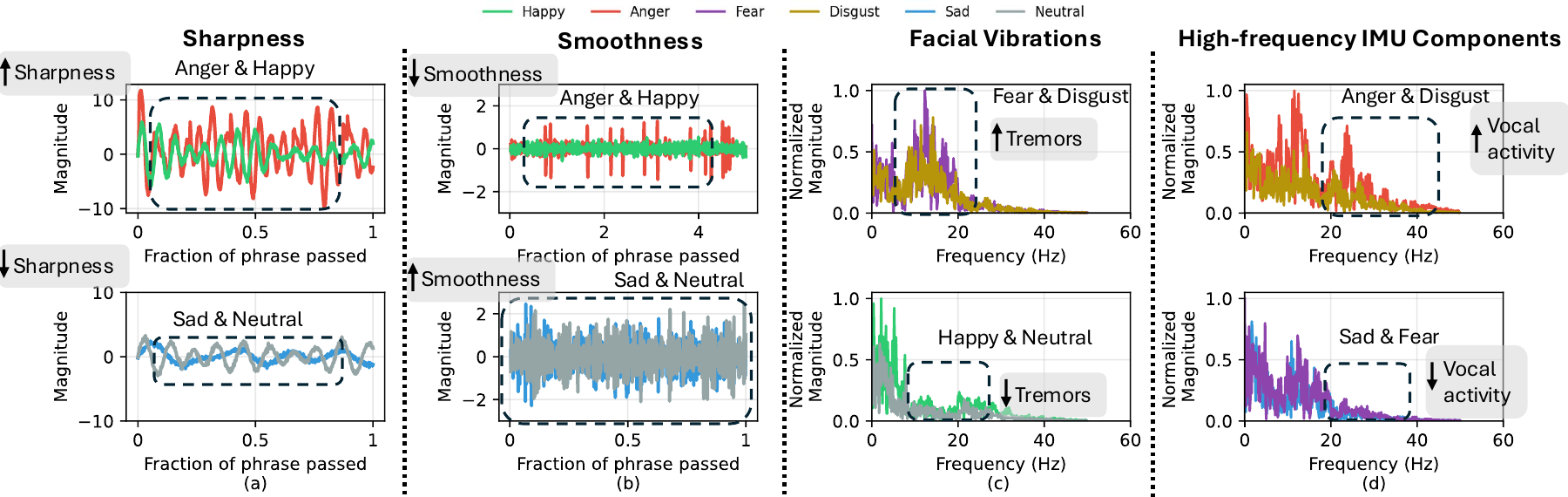}
    \vspace{-0.5cm}
    \caption{Analysis of movement and vibration characteristics during different emotional states, showing (a) sharpness, (b) smoothness, (c) facial vibrations, and (d) high-frequency IMU components across different emotional conditions and frequencies. All features are for the phrase \qq{I cannot believe this.}}
    \label{fig:comb_feat}
    
\end{figure*}

% \vspace{-0.3cm}
\section{Articulatory features}
\label{sec:features}

\system{} combines complementary information from high-fidelity (audio) and sparse modality (IMU).
% Our approach leverages audio for prosodic and spectral features critical for emotion recognition, while IMU provides articulatory signatures absent in audio. 
\textcolor{black}{We observe that articulatory motion patterns vary across different emotional expressions.}
% The articulation of different emotional states produces distinct patterns in the jaw and facial muscle motion.
We isolate the IMU signal into three categories based on frequency.
 % : jaw motion (0-5 Hz), facial vibrations (5-20 Hz), and 
% \textcolor{black}{higher-frequency IMU components (20-50 Hz).}
We discuss the details extracted from each frequency band in the following subsections.

\vspace{-0.3cm}
\subsection{Jaw Motion (0-5 Hz)}
Jaw motion during emotional speech 
\textcolor{black}{exhibits low-frequency patterns associated with different emotional expressions}
~\cite{lee2005articulatory}. For instance, anger 
\textcolor{black}{is often associated with}
forceful and loud articulation with rapid changes in jaw motion, while sadness 
\textcolor{black}{exhibits}
more restrained, measured motion. We extract four key features from jaw motion:

\noindent $\blacksquare$ \textbf{Sharpness:} Sharpness quantifies the abruptness or transition intensity of a motion~\cite{bressem2011rethinking}, often used in gesture or activity recognition. For example, when speaking angrily, jaw motion 
\textcolor{black}{exhibits}
sharp high-amplitude patterns, while sadness 
\textcolor{black}{shows}
gradual and subjugated patterns~\cite{bloch1991specific}. We quantify these characteristics through the sharpness of the motion, calculated as the second derivative of the jaw motion:

{\centering
$ \displaystyle
\begin{aligned}
S_{mov} = \frac{1}{N} \sum_{i=1}^{N-2} |x[i+2] - 2x[i+1] + x[i]|
\end{aligned}
$
\par}

where $x[i]$ represents the gyroscope signal. Figure~\ref{fig:comb_feat}(a) shows 
\textcolor{black}{that}
% high sharpness values indicate
sharp changes are associated with angry emotion, while low values 
\textcolor{black}{are associated with}
% indicate
smooth and gradual moving trajectories of neutral or sad emotion.

\noindent $\blacksquare$ \textbf{Smoothness:} Jaw motion patterns 
\textcolor{black}{tend to show reduced regularity}
during high-arousal emotions like anger and happiness, while they 
\textcolor{black}{may exhibit more regular patterns}
for low-arousal emotions like sad and fear~\cite{ijerph192013440}. We measure these motion irregularities using the length of spectral arc (SPARC), a metric widely used for quantifying smoothness of motion~\cite{scholp2021spectral} and estimate the disturbances in the natural flow of movement. 

{\centering
$ \displaystyle
\begin{aligned}
SPARC = -\int_{0}^{f_c} \sqrt{1 + \left(\frac{d\hat{X}(f)}{df}\right)^2} df
\end{aligned}
$
\par}

where $\hat{X}(f)$ is the normalized Fourier magnitude. Figure~\ref{fig:comb_feat}(b) shows how jaw motion 
\textcolor{black}{maintains relatively higher regularity}
% maintains regularity
during neutral and sad states (higher SPARC values), while 
\textcolor{black}{showing lower regularity}
% exhibiting lower SPARC values
during expressive states of happiness and anger.

\begin{figure}
    \centering
    \includegraphics[width=0.7\linewidth]{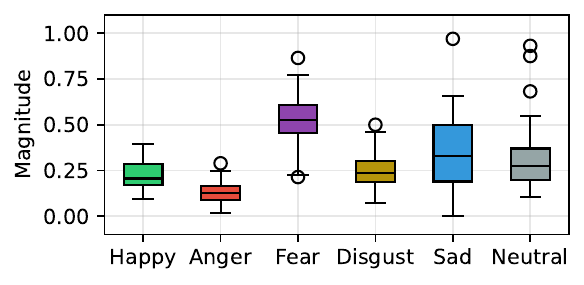}
    \caption{Inter-Burst intervals are comparatively high for low arousal emotions like disgust.}
    \label{fig:feat_temporal_dy}
    \vspace{-0.3cm}
\end{figure}

\noindent $\blacksquare$\textbf{Activity Bursts and Inter Burst Intervals (IBI):} Speech patterns vary between intense articulation and relative stillness, 
\textcolor{black}{and these variations are influenced by emotional expression.}
We analyze the energy hotspots in the signal activity bursts, defined as short periods of intense physical activity~\cite{scherer2013affect}. We calculate the instances in which the magnitude of the gyroscope signal deviates significantly from its mean. In addition to activity bursts, we utilize IBI, which refers to periods of low activity between bursts of activity in jaw motion. We calculate IBI as the time frame for which the gyroscope signal maintains a relatively stable position throughout the phase. Figure~\ref{fig:feat_temporal_dy} illustrates IBI, where sad and neutral states 
\textcolor{black}{tend to show}
longer sustained positions than high-arousal emotions like anger.

\begin{figure}
    \centering
    \includegraphics[width=0.7\linewidth]{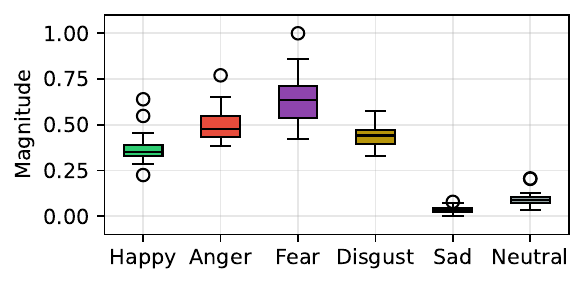}
    \vspace{-0.5cm}
    \caption{Pre-speech build-up rate has higher values for tense emotions like fear and anger.}
    \label{fig:feat_prep_var}
    \vspace{-0.5cm}
\end{figure}

\noindent $\blacksquare$ \textbf{Pre-speech jaw motion features:} A unique aspect of our multimodal IMU-audio system is the analysis of articulatory motions before speech. We find this initial period, $\delta$, as the difference between the onset of IMU and the audio signal. During this preparation phase, we extract two key characteristics: the maximum jaw displacement (\textit{pre-speech amplitude}) and the rate at which this movement builds up (\textit{build-up rate}). These features 
\textcolor{black}{show variations across emotional expressions}
% show rapid and jittery pre-speech characteristics with high amplitude as the speaker tenses up (fear), while anger exhibits sharp, decisive preparation movements. In contrast, sad speech shows gradual, low-amplitude preparation, and happy speech maintains controlled, moderate preparation patterns
~\cite{fuller1992validity}. As shown in Figure~\ref{fig:feat_prep_var}, these distinct preparation signatures 
\textcolor{black}{can help differentiate}
% effectively distinguish
between different emotional states.

\vspace{-0.2cm}
\subsection{Facial Vibrations (5-20 Hz)}
We utilize mid-frequency band features from facial muscle vibrations and skin deformation to extract information about emotional state both before and during speech production. Facial muscles exhibit vibrations that intensify during states of fear or anxiety~\cite{10.1093/chemse/bjad016}. We extract two key features:

\noindent $\blacksquare$ \textbf{Spectral Analysis:} During speech with fear or disgust, our facial muscles exhibit subtle, rapid vibrations that reflect high arousal. We characterize these vibrations from the accelerometer through three key aspects: their overall intensity (\textit{power}), their primary oscillation rate (\textit{dominant frequency}), and how varied these oscillations are (\textit{frequency spread}). Together, these measures characterize muscle tension patterns - fear and disgust show intense, widely distributed tremors across multiple frequencies, indicating unstable muscle activity. On the other hand, happy and neutral exhibit minimal activity with more focused frequency patterns, reflecting stable muscle control. Figure~\ref{fig:comb_feat}(c) shows the frequency spectrum for fear and disgust (top row) with high tremors and happy and neutral (bottom row) with low tremors.

\noindent $\blacksquare$ \textbf{Pre-speech facial vibrations:} Similar to jaw motion, we analyze the facial muscle vibrations in the pre-speech window, computing both pre-speech amplitude and build-up rate. This quantifies how facial muscle activity develops before speech onset, with features such as jitter in vibration intensity showing emotion-specific patterns.
 
\subsection{High-frequency IMU Components (20-50 Hz)}
\textcolor{black}{The high-frequency band captures speech-induced mechanical vibrations transmitted through facial tissues and bone, reflecting motion patterns associated with speech production~\cite{shi2021face}.}
\textcolor{black}{We analyze the energy of this band as an indicator of articulatory activity during speech.}
\textcolor{black}{These IMU components provide complementary motion information to audio signals and are less sensitive to environmental noise compared to air-conducted sound.}
\textcolor{black}{We observe that the energy in this band varies across emotional expressions, with higher values often associated with more expressive or high-arousal speech.}
\textcolor{black}{As shown in Figure~\ref{fig:comb_feat}(d), emotions such as anger and disgust exhibit higher activity in this band, while sad and fear show comparatively lower activity.}

\subsection{Multi-resolution Analysis}

To capture how emotions manifest across different time scales, from rapid muscle activations to slower articulatory patterns, we use wavelet decomposition. This multi-resolution approach helps distinguish emotions that might have similar overall energy but different temporal characteristics (e.g., the quick, chaotic movements of anger versus the suppressed movement of neutral speech). For each decomposition level $l$, we compute:
\begin{equation}
E_l = \sum_{k} |d_l[k]|^2, \qquad
H_l = -\sum_{k} p_l[k] \log p_l[k]
\end{equation}

where $d_l[k]$ represents the detail coefficients and $p_l[k]$ is the normalized energy distribution for $k^{th}$ time index. The wavelet decomposition is applied directly to the raw IMU signals to capture motion patterns at different time scales. The energy $E_l$ and entropy $H_l$ together quantify the intensity and complexity of these multi-scale motion patterns.

\subsection{Statistical Features}
Along with our emotion-related articulatory features, we extract five statistical measures that provide essential baseline information about movement characteristics. We compute standard metrics including mean ($\mu$), standard deviation ($\sigma$), skewness ($\gamma$), and kurtosis ($\kappa$). The crest factor ($CF$):

{\centering
$ \displaystyle
\begin{aligned}
CF = \frac{\max|x[i]|}{RMS}
\end{aligned}
$
\par}

helps quantify movement dynamics - lower values indicate sustained, consistent movements (as in sadness), while higher values suggest sudden, intense bursts (as in anger).

\subsection{Audio Features}
\label{sec:audio_features}
For audio feature extraction, we are using the emobase feature set~\cite{schuller2006emotion} by openSMILE~\cite{eyben2010opensmile}. It has features derived from 26 Low-Level Descriptors (LLDs) and their delta coefficients. The 26 LLDs include important acoustic parameters such as Fundamental Frequency (F0), 12 Mel-Frequency Cepstral Coefficients (MFCCs), Zero-Crossing Rate (ZCR), Probability of Voicing, Intensity, Loudness, F0 Envelope, and 8 Line Spectral Frequencies. This set of features has demonstrated robust performance on a wide range of datasets for the emotion recognition task~\cite{dogdu2022comparison}.

\section{Emotion Recognition Pipeline}
\label{sec:recog}
Our key insight is that emotions have inherent relationships - for example, \qq{fear} and \qq{disgust} are more similar to each other compared to \qq{happy}. We capture these relationships in two ways: first, by using text language model embeddings that encode semantic similarities between emotions, and second, by training our model to predict not just the emotion category, but also its valence (positive/negative) and arousal (intensity) values. This multitask formulation helps our model to recognize not only emotions but also their relationships. 

\begin{figure}
    \centering
    \includegraphics[width=0.8\linewidth]{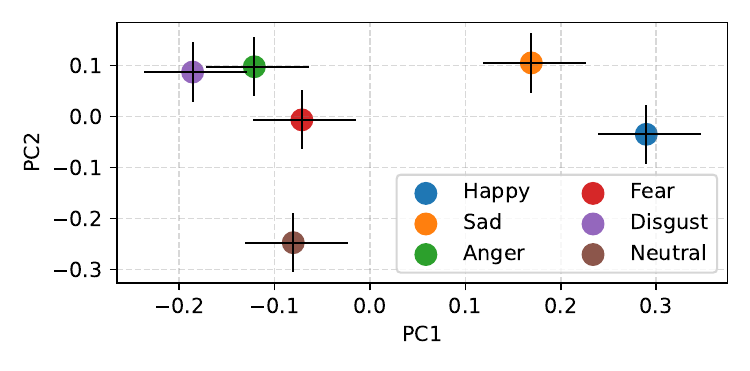}
    \vspace{-0.5cm}
    \caption{Different emotions embedding space from text projected in 2D using PCA. As we can see, emotions with negative valence are close to each other. This behavior is also mimicked by our model's embedding space (Figure~\ref{fig:pca_vis}).}
    \label{fig:text_vis}
   \vspace{-0.5cm}
\end{figure}

\subsection{Learning Emotion Relationships}
We use the embeddings of the RoBERTa~\cite{liu2019RoBERTa} language model 
\textcolor{black}{to encode semantic relationships between emotion labels.}
text has been trained on large amounts of text data, allowing it to learn meaningful representations of words and concepts. As shown in Figure~\ref{fig:text_vis}, it represents different emotion words 
\textcolor{black}{such that semantically related emotions (e.g., \qq{fear} and \qq{disgust}) are closer in embedding space, while contrasting emotions (e.g., \qq{happy} and \qq{sad}) are farther apart.}
Figure~\ref{fig:text_vis} shows the high-dimensional embeddings projected onto a 2D space using principal component analysis, where the x and y axes represent PC1 and PC2, respectively. Due to this dimensionality reduction for visualization purposes, some semantically distant emotions may appear closer in this representation than they are in the original embedding space. 
We specifically use embeddings of single emotion words (e.g., \qq{happy}, \qq{angry}) rather than emotion phrases (e.g., \qq{I am very happy today}, \qq{I am so angry}) since word embeddings 
\textcolor{black}{provide a stable representation of emotion categories.}
To incorporate these relationships into our model, we use a cosine distance-based learning approach:
\begin{equation}
L_{metric} = \sum_{i,j} \max(0, D_{ij} - \alpha_{ij} + m)
\label{eq:metric}
\end{equation}
Here, $D_{ij}$ measures the distance between emotion representations i and j in our model, 
\textcolor{black}{$\alpha_{ij}$ defines a target separation based on semantic similarity,}
and $m$ enforces a margin between them.

\subsection{Multi-task emotion recognition}
Building on the emotion relationships learned through the text embeddings, we extend our model to capture emotional dimensions through the prediction of valence-arousal (VA) as explained in Section~\ref{sec:bg}. Similar to how vision systems benefit from jointly learning related tasks like object detection and segmentation~\cite{zhang2021survey}, emotion recognition can benefit from jointly predicting both categorical emotions and their dimensional attributes. 

\textcolor{black}{We treat VA prediction as an auxiliary task to support emotion classification.}
We define a discrete mapping of emotions into VA space:
\vspace{-0.1cm}
\begin{equation}
    \text{VA}_{\text{mapping}} = 
    \begin{aligned}
        &\{\text{Happy}: (+1,+1), \text{Sad}: (-1,-1), \\
        &\text{Anger}: (-1,+1), \text{Fear}: (-1,+1), \\
        &\text{Neutral}: (0,0), \text{Disgust}: (-1,+1)\}
    \end{aligned}
\end{equation}

While valence and arousal exist on a continuous spectrum, we discretize these values because of the absence of reliable continuous ground-truth annotations for our dataset. 
\textcolor{black}{We do not model continuous VA values, and our mapping is used only to provide coarse supervision for auxiliary learning. This mapping captures broad relationships (e.g., shared arousal levels) and does not represent the full complexity of affective space.}
Our model jointly optimizes emotion classification and VA prediction through a weighted loss:
$L_{total}=\omega_eL_{emotion}+\omega_vL_{valence}+\omega_aL_{arousal}$, where $L_{emotion}$ is the cross-entropy loss for emotion classification, $L_{valence}$ and $L_{arousal}$ are binary classification losses for VA prediction. We set $\omega_e = 0.5$, $\omega_v = \omega_a = 0.25$ to balance these objectives. 
\textcolor{black}{These weights were selected empirically.}
 
The complete framework can be represented as: $f_\theta: X \rightarrow (E, V, A)$, 
where X represents input features, E is the discrete emotion space, and (V,A) represents the valence-arousal predictions. 
This multi-task formulation, combined with the semantic relationships, enables our model to learn both categorical and dimensional aspects of emotions. 

\section{ Data Collection and Implementation}
\label{sec:data}
In this section, we detail our data collection setup, user study, preprocessing pipeline, and model training. 
 \begin{figure}
    \centering
    \includegraphics[width=\linewidth]{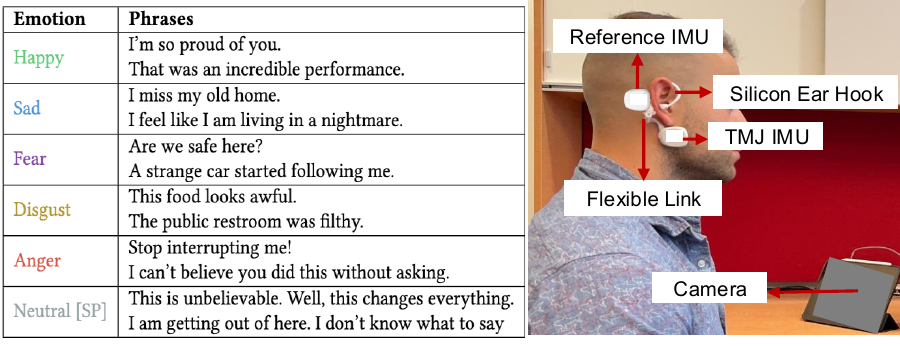}
    \caption{Some example phrases from our user study. We have phrases of length varying from 3 to 8 words. SP - Same Phrases that were common to all the emotions. Our prototype design is inspired by~\cite{srivastava2023jawthenticate, srivastava2024whispering}.}
    \vspace{-0.5cm}
    \label{fig:data_collection_comb}
\end{figure}

\subsection{Data Collection}
We conducted an IRB-approved study with 20 participants (12 male, 8 female) who were fluent in English and one or more additional languages. Data were collected using a twin-IMU setup consisting of a TMJ-mounted IMU, a reference IMU, and a microphone. The IMUs recorded 3-axis accelerometer and gyroscope data at 100 Hz, while audio was sampled at 48 kHz. Participants completed two recording sessions separated by at least three days.

We collected six emotion categories (Happy, Sad, Fear, Disgust, Anger, and Neutral) across six evaluation scenarios: different phrases, same phrases, native-language speech, paragraph reading, conversational responses, and noisy conditions. We obtain emotion labels through participant self-reports following each recording. To assess label consistency, three independent annotators reviewed a subset of the dataset, yielding a Pearson correlation coefficient of 0.93 with participant labels. Detailed data collection procedures, participant demographics, recording protocols, and scenario descriptions are provided in the Appendix~\ref{sec:appendix_dataset}.

\subsection{Signal processing and Model training}
% 
% We use the methodology employed in~\cite{srivastava2023jawthenticate} to remove the effect of motion noise. \sj{Why is this lone sentence hanging out by itself?}
% In our twin-IMU setup, the TMJ IMU captures the jaw motions while users are articulating different phrases from the UI, and the reference IMU captures the body and head motions.
% It does not capture the jaw motions. Another IMU is placed on the temporomandibular joint (TMJ) 
%jaw motions while users are articulating different phrases from the UI. 
Users' body and head movements corrupt the TMJ IMU's signals. 
%\sj{No role of the reference IMU?} 
To attenuate this body motion from the jaw motion, we used a reference IMU (Figure~\ref{fig:data_collection_comb}) placed on the temporal bone of the user and implemented an adaptive finite impulse response (FIR) filter to remove noise~\cite{srivastava2023jawthenticate}.
% , which can return a desired signal modeling the complex and non-linear noise of the input signal. 
To adjust the coefficient of the FIR filter, we used the Least Mean Squares (LMS) algorithm to cancel the noise and obtain optimized performance. In addition, we remove the effect of gravity from the accelerometer data and DC bias using a high-pass filter.

% After segmentation, the clean raw data was subsequently used to initiate the feature extraction process. Our vanilla IMU-audio baseline employed a variety of features, including: \sj{Why are we talking about features for a specific baseline here?} \paras{I agree, not sure to which section I should move this to or remove it completely? any suggestion?}

% \begin{itemize}
%     \item \textbf{Statistical Features:} Mean, Standard Deviation, Skewness, Kurtosis, Root Mean Square (RMS), Maximum, Minimum, Range, Median, Interquartile Range (IQR), Mean Absolute Deviation (MAD)
    
%     \item \textbf{Time Domain Features:} Mean Crossing Rate, Time to Max, Rhythm Mean and Standard Deviation, Smoothness (Spectral Arc Length), Permutation Entropy, Teager-Kaiser Energy, Peak Count, Energy
    
%     \item \textbf{Frequency Domain Features:} Dominant Frequency, Spectral Centroid and Spread, Spectral Entropy
    
%     \item \textbf{Time Series Features:} Autocorrelation, Trend
    
%     \item \textbf{Non-linear Features:} Sample Entropy, Largest Lyapunov Exponent
    
%     \item \textbf{Autoregressive Features:} AR coefficients (up to order 5)
% \end{itemize}

% \noindent $\blacksquare$ {\bf Model Training.}
After extracting the features described in Section~\ref{sec:features}, we train a classifier for emotion recognition. We experimented with multiple models
% , including Support Vector Machine (SVM), Naive Bayes, Random Forest, Neural Network, Nearest Neighbor, and XGBoost. To our findings, 
and found XGBoost to perform the best for the baseline features, and an ensemble model
% . While for \system{} features the best performance was observed with an Ensemble model 
comprising of Support Vector Machine (SVM), Neural Network, and Nearest Neighbor models best for \system{}. 
% We take a weighted sum of the outputs of each of the models in the Ensemble model. We also tuned the hyperparameters for all these models. 
For our evaluation, we conducted a rigorous leave-one-user-out cross-validation protocol to assess the generalizability of our approach. For each test user, we additionally used 20\% of their stratified data for fine-tuning the system, which helped adapt the model to individual differences in emotion expression.

% \noindent $\blacksquare$ \textbf{\paras{system_name} Model Training:}We experimented with multiple models, including Support Vector Machine (SVM), Naive Bayes, Random Forest, Neural Network, and XGBoost. To our findings XGBoost performed the best \sj{Then why did you combine it with the other 2?}, followed by SVM and Neural Network. We then tuned the hyperparameters for these models
% \begin{itemize}
%     \item For \textbf{Support Vector Machine (SVM)}, we tuned the 'C' and 'gamma' hyperparameters
    
%     \item For \textbf{Neural Network}, we optimized the number of neurons and epochs
    
%     \item For \textbf{XGBoost}, we used a parameter grid to tune 'max\_depth', 'learning\_rate', 'n\_estimators', 'min\_child\_weight', 'subsample', 'colsample\_bytree', and 'gamma'.

% \end{itemize}

\begin{figure*}[t]
    \centering
    \includegraphics[width=\linewidth]{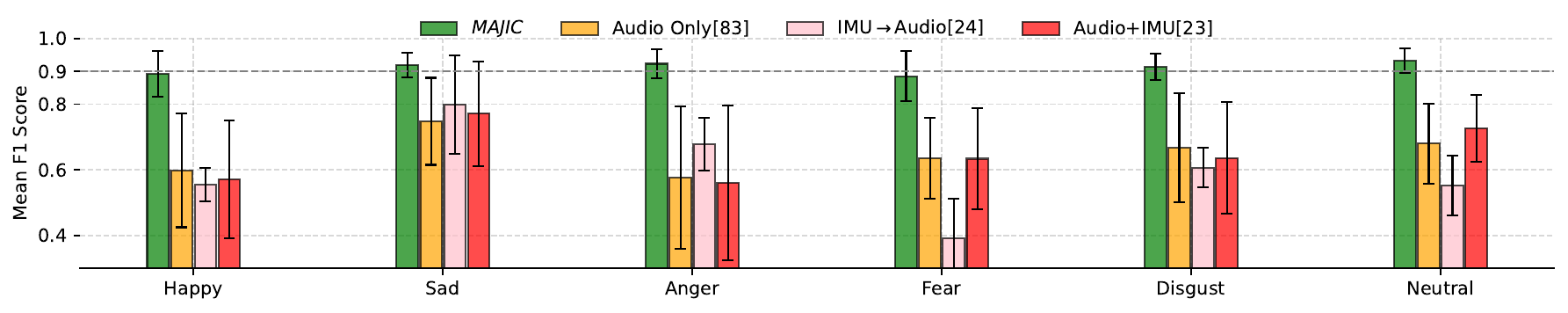}
    \vspace{-0.5cm}
    \caption{The mean F1-Score for our system is greater than 90\%, surpassing the other baselines by at least 26.2\% on average.}
    \label{fig:main_res}
\end{figure*}

% \noindent $\blacksquare$ {\bf Baselines}
\subsection{Baselines and Metrics}
We compare \system{} with 3 baselines implemented from state-of-the-art solutions in audio and jaw motion. For all baselines, we train a personalized model for each user as described above. We follow the same model architecture and training above for the baselines and pick the best-performing hyperparameters. 
% We use the latest audio-based system, Unvoiced, and the Jawthenticate (IMU) + audio-based system for a baseline comparison with our system. 

% by splitting data by leave-one-user-out and data 20\% to fine-tune the model from the target user. We trained an XGBoost model for all baselines, as it performed the best, stated otherwise \sj{again, what? So you're using ensemble and everyone else uses XGBoost?}. 

%To evaluate improvements over audio-only approaches, we implemented two audio baselines: (1) We fine-tuned emotion2vec~\cite{ma2024emotion2vec} on our dataset, which is an emotion representation model for extracting speech features for emotion recognition. (2) We utilized the features described in Section~\ref{sec:features} and trained various classifiers as detailed in our Model Training section. These features align with those used extensively in recent SER research ~\cite{Kyaw2024XGBoostbasedME, dogdu2022comparison} Since baseline (2) shows better performance compared to emotion2vec, we present results only for this better-performing audio-based classifier.

\noindent \emph{$\bullet$Audio-only Baseline}:  We evaluated state-of-the-art pretrained SER models, specifically Emo2Vec~\cite{ma2024emotion2vec}, against our emobase feature extraction approach (Section~\ref{sec:features}). Emo2Vec was fine-tuned on our dataset using the same user-specific adaptation protocol employed in our system (20\% stratified data per user). However, as the emobase feature set performs better than Emo2Vec, we use emobase as our audio baseline. We discuss the results in Section~\ref{sec:main}.

\noindent\emph{$\bullet$IMU$\rightarrow$Audio\cite{srivastava2024unvoiced}}: As a second audio-only baseline, we used Unvoiced~\cite{srivastava2024unvoiced} to transform the speech-related jaw motion into a rich representation of speech, the mel spectrogram. We used this IMU-generated spectrogram to extract audio features as the previous baseline. 
%$\sj{We should probably rename this to maybe something like IMU$\rightarrow$Audio? Audio seems misleading.}
% train a CNN classifier for this baseline.

\noindent \emph{$\bullet$Audio + IMU\cite{srivastava2023jawthenticate}:} In this baseline we combine audio features with IMU-derived features described in Jawthenticate~\cite{srivastava2023jawthenticate}. 
% We used Jawthenticate~\cite{srivastava2023jawthenticate} as one of our baselines as it 
Jawthenticate extracts characteristics such as intonation and rhythm that are indicative of the emotion expressed in the phrase. In addition to these features, we also include statistical features from the IMU.
% feature set and including emobase audio features from the audio. 

% \noindent $\blacksquare$ 
\textbf{Metrics:} We evaluate our system using standard performance metrics widely adopted in the literature on emotion recognition: accuracy and F1 score \cite{cheng2024emotion, kayande2023fine}. 
% Accuracy measures the overall classification performance for all emotions. F1 score, the harmonic mean of precision and recall, gives a wider view of performance. 
Together, these metrics comprehensively assess our system's emotion recognition capabilities across different scenarios.

\section{Evaluation}
In this section, we show that ~\system{}: 
\textcolor{black}{(1) achieves strong performance on our dataset with an accuracy of 93\% and a 91\% F1 score for emotion recognition using articulatory motion and audio features; (2) shows consistent performance across users and languages with limited training data, (3) distinguishes emotions expressed using the same phrase, and (4) maintains performance across multiple data collection scenarios including conversational prompts.}

\subsection{Overall System Performance}
\label{sec:main}

\begin{figure*}[th]
\centering
\begin{minipage}{0.36\textwidth}
    \includegraphics[width=0.8\linewidth]{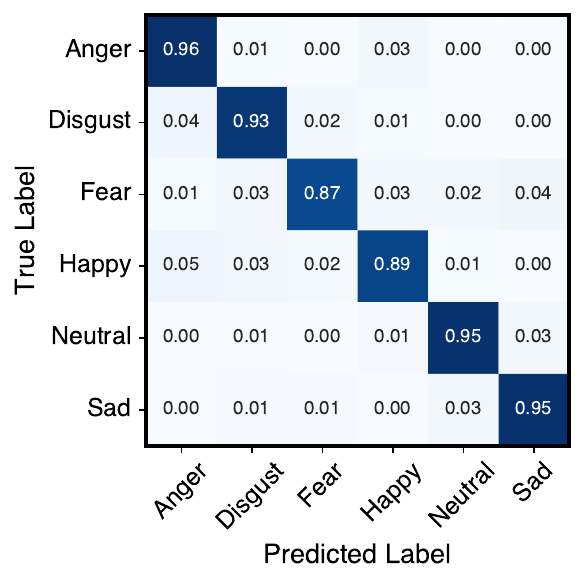}
    \vspace{-0.3cm}
    \caption{\small{Combined confusion matrix for all the users. }}
    \label{fig:conf_mat}
\end{minipage}%
\hfill%
\begin{minipage}{0.30\textwidth}
    \includegraphics[width=0.8\linewidth]{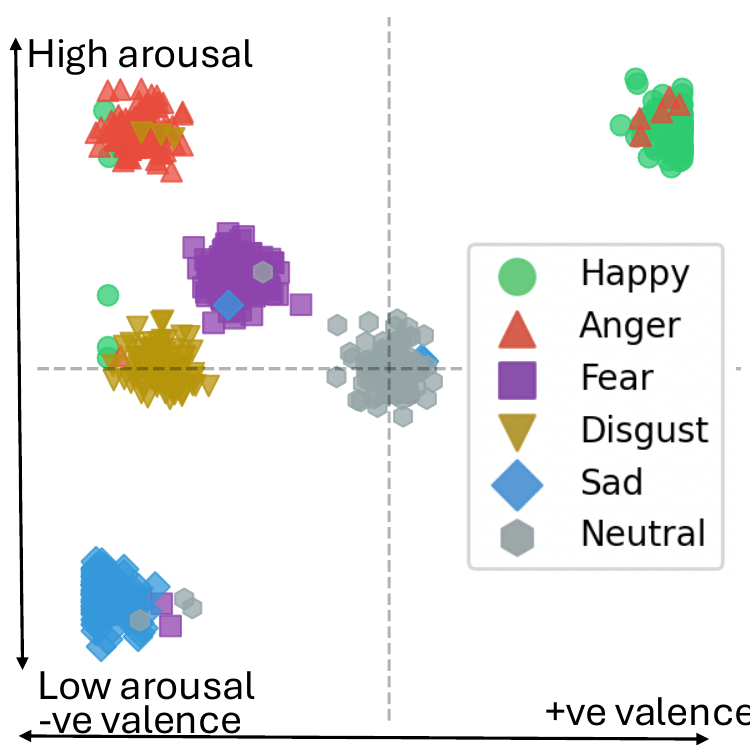}
    \caption{\small{Relative distance between the clusters of different emotions formed using PCA on our feature set and projected on the Valence and Arousal axis using~\system{}'s VA predictions.}}
    \label{fig:pca_vis}
\end{minipage}%
\hfill%
\begin{minipage}{0.30\textwidth}
    \includegraphics[width=0.8\linewidth]{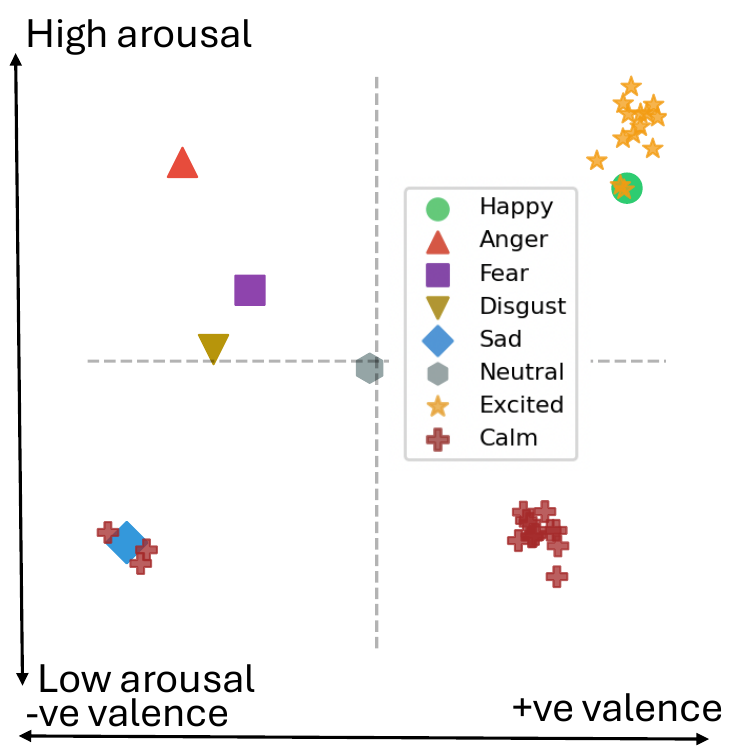}
    \caption{\small{Adding new emotions \qq{Excited} and \qq{Calm} into our valence-arousal space. The centroids of the other six emotions are extracted from Figure~\ref{fig:pca_vis}.}}
    \label{fig:new_emo}
\end{minipage}%
\end{figure*}

\noindent $\blacksquare$ \textbf{Emotion Classification:} Figure~\ref{fig:conf_mat} shows the confusion matrix for emotion recognition. 
Our system achieves high accuracy for emotions: anger (96\%) and sadness (95\%). Interestingly, confusion occurs primarily between emotions with similar arousal levels but different valence. For example, happiness and anger (both high arousal) show some confusion (3\%), but our system successfully distinguishes them in 89\% of cases by using their distinct jaw motion patterns. This performance is further validated by Figure~\ref{fig:pca_vis}, which shows how~\system{} clusters emotions in the valence-arousal space, aligning with the traditional circumplex model. The separation between high-arousal (anger, happiness) and low-arousal clusters (sadness, neutral) shows our feature set's effectiveness in capturing emotion-specific characteristics.

\noindent $\blacksquare$ \textbf{Comparison with baselines:} Our system achieves a mean F1 Score of 91\%, significantly outperforming baselines as shown in Figure~\ref{fig:main_res}.
For the audio-only baseline, we evaluated Emo2Vec and emobase features. Emo2Vec achieved a 65.2\% F1 score compared to the emobase approach’s 72.1\% performance. \textcolor{black}{These baselines represent both pretrained model-based approaches (Emo2Vec) and traditional feature-based methods (emobase).} \textcolor{black}{We observe that pretrained models trained on large-scale datasets may not directly generalize to our dataset, while feature-based approaches capture dataset-specific variations.}
The emobase features~\cite{dogdu2022comparison}, extracted directly from our non-actor dataset, better capture the subtle emotional variations that characterize everyday speech patterns that pretrained models are not optimized for. \textcolor{black}{The emobase baseline (72.1\%) shows difficulty distinguishing certain emotion pairs.} For example, both happiness and anger may exhibit an increase in volume and pitch, but their jaw motion patterns differ. We implement a baseline with only statistical IMU features and train it in the same way as other baselines. It achieves 62.8\% F1 Score, showing the need for better representative features for IMU; we therefore omitted this baseline from the figure for clarity. Audio+IMU\cite{srivastava2023jawthenticate} is the best-performing baseline with 74.3\% mean F1 score, as it combines IMU characteristics such as intonation and rhythm with statistical IMU and audio.

\begin{figure}
    \centering
    \includegraphics[width=1\linewidth]{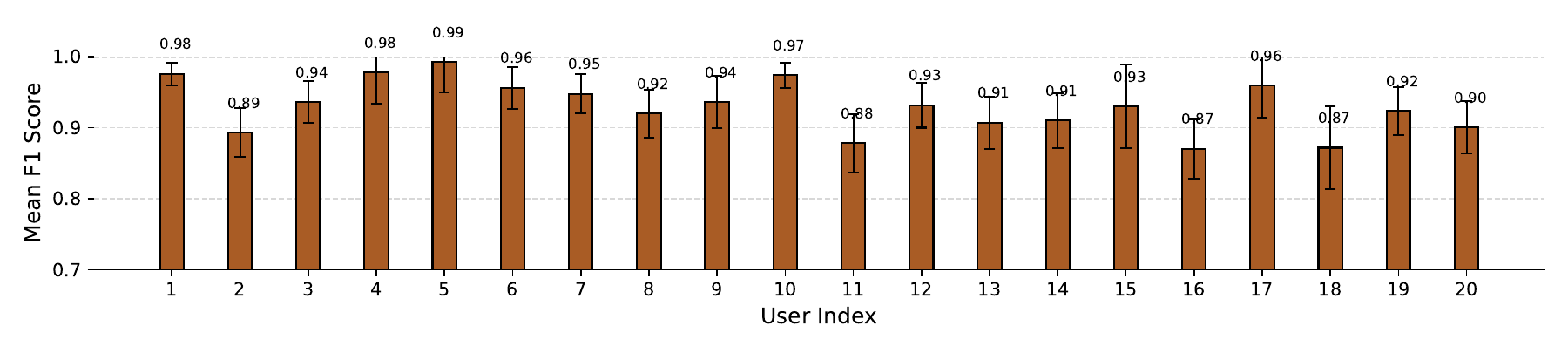}
    \vspace{-0.8cm}
    \caption{We achieve $>$85\% F1 score for all of our users.}
    \label{fig:user_res}
    \vspace{-0.5cm}
\end{figure}

\noindent $\blacksquare$ \textbf{User-specific Evaluation:} 
Figure~\ref{fig:user_res} shows the accuracy for 20 users ranges from 87\% to 99\%, demonstrating the robustness of ~\system{} to our diverse participant pool with varying accents, speaking styles, and articulatory patterns. The system maintains this performance for multiple sessions per user, indicating robustness to variations in sensor placement and natural changes in speech patterns. Please refer to the Appendix for per-user accuracy.

\begin{figure*}[th]
\centering
\begin{minipage}{0.25\textwidth}
\vspace{0pt}
\includegraphics[width=\linewidth]{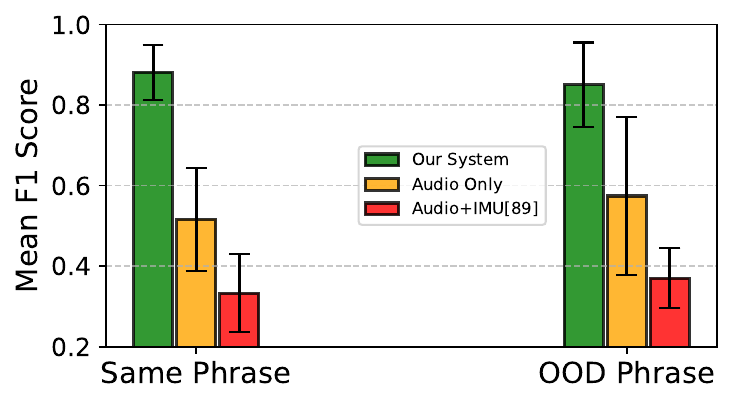}
    % \vspace{-18pt}
    \caption{\small{Our system performs well in two challenging scenarios: (1) Same phrase, and (2) Out-of-dictionary phrases.}}
    \label{fig:phase_res}
\end{minipage}%
\hfill%
\begin{minipage}{0.4\textwidth}
\vspace{-8pt}
\includegraphics[width=\linewidth]{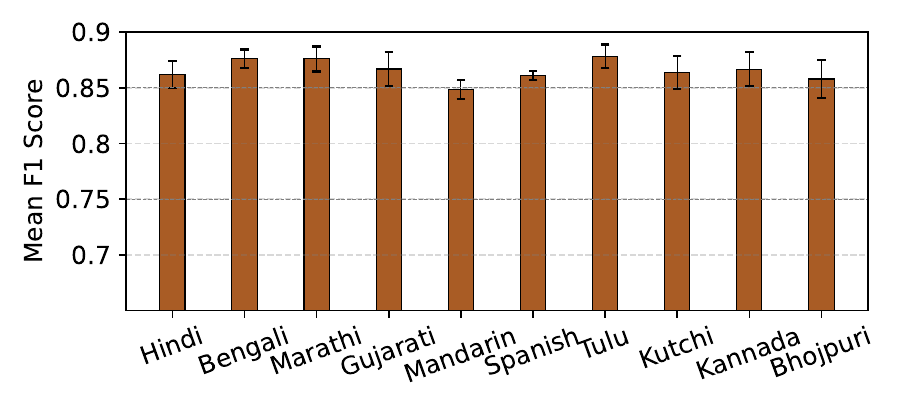}
    % \vspace{-18pt}
    \caption{\small{Performance of our system when the training set is in English, and the testing set is in the user’s native language.  }}
    \label{fig:lang_res}
\end{minipage}%
\hfill%
\begin{minipage}{0.29\textwidth}
% \vspace{-8pt}
    \includegraphics[width=\linewidth]{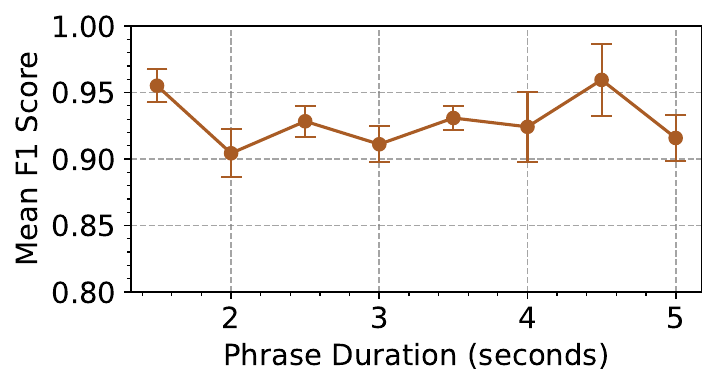}
    \vspace{-18pt}
    \caption{\small{Our system's performance stays consistent with phrases of varying lengths from 1.5 to 5 seconds.}}
    \label{fig:phrase_len_res}
\end{minipage}%
\hfill%
\end{figure*}

\subsection{Content and Language Invariance}

\noindent $\blacksquare$ \textbf{Same phrase:}
One of the main challenges in SER is maintaining accuracy when users express the same phrase through different emotions. For example, \textit{\qq{Oh, my god}} for happy and sad emotions. Our system achieves $>$90\% F1 score in such scenarios (Figure~\ref{fig:phase_res}), 
\textcolor{black}{indicating that articulatory features capture variations in emotional expression even when spoken content remains unchanged.} Moreover, for out-of-dictionary (OOD) phrases, we maintain $>$85\% F1 Score, 
\textcolor{black}{suggesting generalization to unseen phrases within our dataset.}(Figure~\ref{fig:phase_res}).

\noindent $\blacksquare$ \textbf{Cross-language performance:}
Training in English data and testing in nine other languages, we achieve an accuracy exceeding 80\% in all cases (Figure~\ref{fig:lang_res}). 
\textcolor{black}{This performance suggests that the system captures articulatory and acoustic features that are less dependent on specific linguistic content.}
Interestingly, we observe slightly higher accuracy (85\%) for languages with prosodic characteristics similar to English, while maintaining 
\textcolor{black}{comparable performance (80\%) even for tonally distinct languages such as Mandarin. These results indicate that \system{} generalizes across the evaluated languages.}

\noindent $\blacksquare$ \textbf{Robustness to content length:}
We evaluate \system{} with phrases of varying lengths. As shown in Figure~\ref{fig:phrase_len_res}, we maintain a $>$90\% F1 score in phrases ranging from 1.5 to 5 seconds. This stability is critical for real-world applications where utterance lengths naturally vary. We observe an improvement in accuracy for longer phrases, likely due to the availability of more emotional context, but even with phrases that last only 2 seconds, we achieve an F1 score of 91\%.

\begin{figure}
    \centering
    \includegraphics[width=0.8\linewidth]{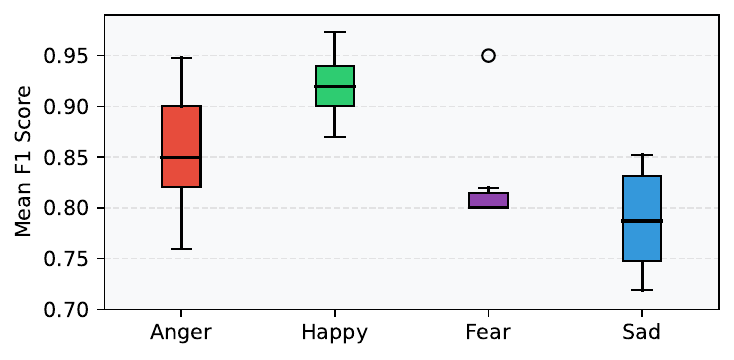}
    \caption{Recognition accuracy across four emotions for conversational data.}
    \label{fig:conv_res}
    \vspace{-0.3cm}
\end{figure}

\noindent $\blacksquare$ \textbf{Conversations:}
To evaluate the performance of our system  \textcolor{black}{on conversational data collected using prompts,}
we compare the detected emotion with self-reported emotion for 2-second windows. 
We observed that users rarely maintain consistent emotional expression throughout an entire conversation, and 2 seconds was empirically observed to work better than other segment lengths. 
As shown in Figure~\ref{fig:conv_res}, we observe higher recognition accuracy for high-arousal emotions (Happy: 90\%, Anger: 85\%) compared to more subtle emotional states (Sad: 79\%). 
\textcolor{black}{This trend is consistent with the observation that high-arousal emotions often exhibit stronger acoustic and articulatory cues.}
\textcolor{black}{Overall, these results indicate that the system can capture emotional variations in conversational-style data.}

\noindent $\blacksquare$ \textbf{Paragraphs:} 
We observe variation in the mean F1 score for emotion recognition across paragraph segments (2s each). 
High-arousal emotions (anger, happiness) 
\textcolor{black}{tend to show higher recognition accuracy}, suggesting they are easier to recognize when expression intensity is higher. Low-arousal emotions like disgust  \textcolor{black}{show more gradual changes in performance} reflecting more subtle expression patterns.
% \textcolor{black}{The reduced performance for some emotions may reflect the difficulty of consistently expressing subtle emotions over longer durations.} \textcolor{black}{Overall, these results indicate the system captures temporal variations in emotional expression.}

\subsection{Adding a new user}
\label{sec:add_new_user}
\begin{figure}
    \centering
    \includegraphics[width=\linewidth]{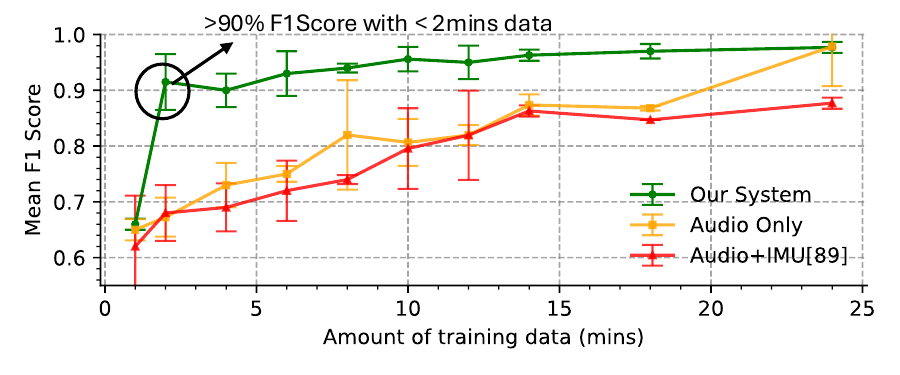}
    \vspace{-0.5cm}
    \caption{While other baselines need $>$ 20 minutes of training data to we can achieve $>90$\% F1 Score, with ~2 minutes of training data.}
    \label{fig:training_requirement_res}
    \vspace{-0.2cm}
\end{figure}

To scale the system to new users, our goal is to minimize the data collection burden for new users while accommodating various cultural biases. As shown in Figure~\ref{fig:training_requirement_res}, our system achieves $>$90\% F1 score with only 2 minutes of training data per user, significantly reducing the effort in real-world deployments. In contrast, baseline approaches require much more data to achieve similar performance. The audio-only baseline requires 24 minutes of speech data to achieve $>$90\% F1 score. Despite the ease of data collection, asking users to perform continuous emotional expressions for extended periods is physically and mentally taxing.

\subsection{Recognizing unseen emotion}
\label{sec:add_new_emotion}

\textcolor{black}{We explore whether the learned representations extend to emotions not explicitly included in the training set.} \textcolor{black}{As an exploratory analysis,} we test \qq{Excited} and \qq{Calm} as additional emotion categories. For each emotion, we collected data from 3 participants, 5 phrases per emotion. 
We predicted the valence and arousal values for these test emotions using our model trained on the six basic emotions in Figure~\ref{fig:pca_vis}. Figure~\ref{fig:new_emo} shows how these test emotions map onto our model's valence-arousal space. 
\textcolor{black}{We observe that}
\qq{excited} shows high valence and high arousal (upper-right quadrant), while \qq{calm} shows positive valence with low arousal (lower-right quadrant). 
The centroids of the six basic emotions serve as reference points for evaluating these new categories. 
\textcolor{black}{We observe limited confusion with nearby emotion classes (20\% of Excited → Happy and 30\% of Calm → Sad).}
\textcolor{black}{These observations are consistent with expected relationships in valence-arousal space.}

\subsection{Sensitivity to noise}
\noindent $\blacksquare$ \textbf{Impact of sensor placement:}
To assess the robustness of ~\system{} to variations in sensor placement, we collect data across 2 sessions and perform a leave-one-session-out evaluation. 
% During the first session, users were guided on the correct placement of sensors, while in the subsequent session, they independently placed the sensors. As a result, 
Sensor positions varied across sessions since users mounted their own prototype. 
% These sessions were held on different days, with gaps of more than a week. 
For the leave-one-session-out evaluation, the model is trained on data from one session and tested on the other session. The mean F1 score exceeds 85\%, with less than 7\%  mean variation between sessions. 
% These results demonstrate that the system maintains reliable performance despite minor inconsistencies in sensor placement between sessions.

\noindent $\blacksquare$ \textbf{Impact of motion and acoustic noise:} We evaluated the robustness of our system to body motion artifacts by training the model with data collected when users were still and testing it with data collected during walking movement. We train in a leave-one-user-out manner and fine-tune with 20\% data from the test user. Our system maintains a high average F1 Score of 88\% ($\pm$0.05) across all users during walking, with scores ranging from 85\% to 91\%. When evaluated in the presence of acoustic noise, we achieve an F1 score of 86\%, 85\%, and 84\% among all users for restaurant, rock music, and hallway background noise, respectively.

\subsection{Ablation Study}
\label{sec:abla_sec}

\begin{table}[t]
\centering
\setlength{\tabcolsep}{0pt} % Remove default padding between columns
\begin{tabular}{|p{1.2cm}|p{1.2cm}|p{1.2cm}|p{1.2cm}|p{1.2cm}|p{1.2cm}|p{1.2cm}|}
\hline
\centering\textbf{F1 Score} & \centering\textbf{Audio} & \centering\textbf{JM} & \centering\textbf{FV} & \centering\textbf{HF IMU} & \centering\textbf{Sem Embed} & \centering\textbf{Multi-task} \tabularnewline \hline
\centering \textbf{91\%} & \cellcolor[HTML]{4CAF50} & \cellcolor[HTML]{4CAF50} & \cellcolor[HTML]{4CAF50} & \cellcolor[HTML]{4CAF50} & \cellcolor[HTML]{4CAF50} & \cellcolor[HTML]{4CAF50} \tabularnewline \hline
\centering \textcolor{red}{71\%} &  & \cellcolor[HTML]{4CAF50} & \cellcolor[HTML]{4CAF50} & \cellcolor[HTML]{4CAF50} & \cellcolor[HTML]{4CAF50} & \cellcolor[HTML]{4CAF50} \tabularnewline \hline
\centering \textcolor{red}{63\%} & \cellcolor[HTML]{4CAF50} &  & \cellcolor[HTML]{4CAF50} & \cellcolor[HTML]{4CAF50} & \cellcolor[HTML]{4CAF50} & \cellcolor[HTML]{4CAF50} \tabularnewline \hline
\centering \textcolor{red}{65\%} & \cellcolor[HTML]{4CAF50} & \cellcolor[HTML]{4CAF50} &  & \cellcolor[HTML]{4CAF50} & \cellcolor[HTML]{4CAF50} & \cellcolor[HTML]{4CAF50} \tabularnewline \hline
\centering \textcolor{red}{73\%} & \cellcolor[HTML]{4CAF50} & \cellcolor[HTML]{4CAF50} & \cellcolor[HTML]{4CAF50} &  & \cellcolor[HTML]{4CAF50} & \cellcolor[HTML]{4CAF50} \tabularnewline \hline
\centering \textcolor{red}{79\%} & \cellcolor[HTML]{4CAF50} & \cellcolor[HTML]{4CAF50} & \cellcolor[HTML]{4CAF50} & \cellcolor[HTML]{4CAF50} &  & \cellcolor[HTML]{4CAF50} \tabularnewline \hline
\centering \textcolor{red}{82\%} & \cellcolor[HTML]{4CAF50} & \cellcolor[HTML]{4CAF50} & \cellcolor[HTML]{4CAF50} & \cellcolor[HTML]{4CAF50} & \cellcolor[HTML]{4CAF50} &  \tabularnewline \hline
\end{tabular}
\caption{Ablation study results showing the performance impact of removing individual components. Green cells indicate components present. JM: Jaw motion, and FV: Facial and HF IMU: High-Frequency IMU}
\label{tab:ablation_res}
\vspace{-1cm}
\end{table}

We perform ablation experiments to quantitatively assess the impact of individual components on ~\system{}'s performance. By removing each component at a time from the system, we report the F1 score in Table~\ref{tab:ablation_res}. \textcolor{black}{We observe performance reductions when individual components are removed, indicating that each modality contributes to the overall system.} \textcolor{black}{Removing audio features results in a drop to 71\% F1 score, suggesting that audio provides complementary information to IMU features.}
Removing jaw motion, facial vibrations, and \textcolor{black}{higher-frequency IMU components}
causes performance drops of 30\%, 28\%, and 20\%, respectively. \textcolor{black}{These results indicate that different IMU components capture complementary aspects of articulatory motion.} \textcolor{black}{Similarly, removing semantic embeddings and the multi-task objective results in performance drops of 13\% and 10\%, respectively, suggesting their contribution to the learned representation.} \textcolor{black}{Notably, removing jaw motion results in a 63\% F1 score, which is lower than the audio-only baseline (72.1\%). This indicates IMU features alone are not sufficient and are most effective when combined with audio.}

\section{Related Work}
% Emotion recognition is widely used in affective computing, healthcare care, and human-computer interaction (HCI), gaining much popularity \cite{erat2024emotion, garcia2024systematic}. 
Many techniques have been developed in affective computing to categorize human emotional states. Table~\ref{tab:comparison} provides a comparison of relevant speech-based emotion recognition systems with our proposed system. 
% In the following paragraphs, we will discuss existing works on emotion recognition based on the modality used.
% using various methods such as speech-based \sj{are we using speech-based or audio-based throughout the paper?}\benjir{we used audio-based, used speech here to refer to audio, should I remove term speech?}, text-based, single-mode, and multi-modal. 

\noindent \textbf{Audio-based Methods.}
SER uses representative features such as tone and pitch
% such as zero crossing rate, root mean square deviation (RMSE), Mel frequency cepstral coefficients (MFCC), pitch, short-term energy (STE), etc. 
extracted from audio for emotion recognition~\cite{deshmukh2019speech, vaidehi2024machine}. These features capture varied characteristics of emotional states in speech. For example, sadness tends to have a lower pitch and energy; on the other hand, anger and excitement have a higher pitch and energy. 
Deep learning techniques are applied to these features to classify emotions \cite{singh2023speech, kalra2023lstm, ottoni2023deep}. These techniques are tested on benchmark datasets such as RAVDESS \cite{livingstone2018ryerson}, TESS \cite{dupuis2010toronto}, EMODB, and SAVEE \cite{haq2009speaker}. These datasets consist of speech data from professional actors expressing several emotional states.
% and serve as benchmark datasets for speech-based emotion recognition, 
% achieving accuracy exceeding 80\%. \cite{ben2024enhancing, tyagi2024exploring, pujitha2024enhanced}. 
% Should I keep this line?% 
However, even SER works, achieving accuracy exceeding 80\%~\cite{ben2024enhancing, tyagi2024exploring}
% the performance drops in real-world scenarios due to factors such as variability in accents, languages, and 
% these solutions 
are hard to translate to non-actors and often suffer in the presence of background noise \cite{zaragoza2024emotional, deschamps2021end}. To address this, our system complements audio features with jaw motion-based features for robust emotion recognition.

\noindent \textbf{\textcolor{black}{Multi-Modal Methods.}} Audio-video systems have been widely used for emotion recognition, leveraging popular datasets: RAVDESS, CREMA-D, CMU-MOSEI, and IEMOCAP, typically combining CNNs and LSTMs for multimodal feature fusion \cite{cao2014crema, zadeh2018mosei}. These systems suffer from occlusion issues, lighting variations, and privacy concerns in everyday deployment \cite{Kjeldsen52:online}. While Sharafi et al.~\cite{sharafi2023audio} achieve $>$95\% recognition accuracy for SAVEE, they utilize audio, 3D dynamic video, and facial markers painted on actors' faces. Text-based augmentation methods extract semantic features from speech transcripts to complement acoustic analysis, though these approaches remain dependent on linguistic content \cite{acheampong2020text, alswaidan2020survey}. These systems struggle with non-verbal emotion cues like sarcasm or tone-dependent emotions, and have cross-linguistic performance issues~\cite{eke2020sarcasm}. In contrast, ~\system{} can recognize different emotions expressed via the same phrase (sarcasm) and shows high accuracy for multiple languages. More recent approaches have incorporated diverse physiological signals: EMG for muscle activity, ECG for heart rate variability, EEG for neural patterns, photoplethysmography (PPG), galvanic skin response (GSR), body posture analysis, and respiratory patterns to capture the broader physiological manifestations of emotional states \cite{fu2022emotion, chen2022emotion, castellano2007multimodal}. Unlike~\system{}, these systems require a multi-sensor setup, place multiple sensors on cheeks and lips, and have not been evaluated in the presence of motion and other noisy conditions. 

Articulatory motion has demonstrated value for emotion recognition, with electromagnetic articulography (EMA) systems showing that jaw opening degree increases with annoyance and lateral lip distance correlates with emotional expression \cite{erickson1998articulatory, nordstrand2004measurements}. Shah et al. achieve high emotion recognition for the non-actor population~\cite{shah2019articulation}. However, they use sophisticated EMA sensors with invasive sensor placement, on the tongue tip, the lower maxilla (for the jaw movement), and the lower lip~\cite{lee2005articulatory}. On the other hand, we achieve competitive accuracy using only a low-cost earable IMU system. Without relying on specialized lab equipment, sensor array, or high-resolution 3D cameras, ~\system{} leverages IMUs that can be integrated into consumer earable devices and achieves high emotion recognition performance for everyday users.

\begin{table}[!t]
\centering
\resizebox{0.5\textwidth}{!}{%
\begin{tabular}{|l|c|c|c|c|c|c|}
\hline
\textbf{System} & \textbf{Modality} & \textbf{\makecell{AF}} & \textbf{\makecell{VF}} & \textbf{\makecell{$>$1L}} & \textbf{\makecell{NA}} & \textbf{Accuracy} \\ \hline

% \cite{brady2016multi} & \makecell{A+V+ECG+EDA+ \\ HRHRV+SCL+SCR} & \ding{55}                                   & \ding{51}                                   & \ding{55}                                 & \ding{51}         \\ \hline

\cite{choudhary2022speech} & A & \ding{55}                                   & \ding{51}                                   & \ding{55}                                 & \ding{55}   & 84.65\%          \\ \hline

\cite{bhattacharya2022emotion} & A & \ding{55}                                   & \ding{51}                                   & \ding{51}                                 & \ding{55}   & 71.43\%          \\ \hline

\cite{kadam2024speech} & A & \ding{55}                                   & \ding{51}                                   & \ding{55}                                 & \ding{55}         & 80.44\%    \\ \hline

\cite{sharafi2023audio} & A+V & \ding{55}                                   & \ding{51}                                   & \ding{55}                                 & \ding{55}    & 95.48\%      \\ \hline

\cite{wang2022multi} & A+V & \ding{51}                                   & \ding{51}                                   & \ding{55}                                 & \ding{55}  & 87.89\%         \\ \hline

\cite{mittal2020m3er} & A+V+T & \ding{51}                                   & \ding{51}                                   & \ding{55}                                 & \ding{55}  & 89.00\%         \\ \hline

\cite{wang2022multi} & A+EEG & \ding{51}                                   & \ding{51}                                   & \ding{55}                                 & \ding{51}  & 89.70\%         \\ \hline

\cite{shah2019articulation} & A+EMA & \ding{51}                                   & \ding{51}                                   & \ding{55}                                 & \ding{51}  & 97.05\%         \\ \hline

\textbf{\system{}} & \textbf{IMU+A} & \ding{51}                                   & \ding{51}                                   & \ding{51}                                 & \ding{51}     & 93.00\%    \\ \hline
\end{tabular}%
}
\caption{Comparison of~\system{} with other SER systems. AF: Articulatory features; VF: Vocal Features; $>$1L: validated on multiple languages; NA: Non-actor dataset. A: Audio and V: video.} % Table title at the bottom
\label{tab:comparison}
\vspace{-1cm}
\end{table}

\section{Discussion}

% \subsection{Discussion}

\noindent \textbf{Exit User Survey Findings.}
We conducted an anonymous exit survey to get user feedback on our system and the data collection process. The survey used a Likert scale to assess system comfort, ease of emotion expression, ability to maintain emotions during paragraphs, and transitions between emotional states. Users reported varying levels of difficulty in enacting different emotions. While most users were comfortable expressing happy, sad, disgust, and neutral, many (60\%) users found it challenging to express fear without external stimuli, having to rely on imagination. 
% In the future, this can be investigated further by narrowing the scope of emotions and collecting data in more realistic settings - such as conversations.
% However, our system's ability to capture secondary articulators (jaw motions and facial vibrations) helped distinguish fear from other emotions even when vocal modulation was limited.

\noindent \textbf{Emotion in longer speech sequences.}
Paragraph-based data collection revealed interesting patterns in the duration of emotion expression. Some emotions, like happiness and anger, benefited from contextual build-up, with users gradually intensifying their expression. In contrast, emotions such as sadness and fear are manifested as sudden bursts. This aligns with natural emotion expression patterns, where some emotions evolve gradually while others appear more abruptly.

%\noindent \textbf{Accuracy based on gender.} There is evidence in the literature pointing to disparities between emotion recognition performance for male versus female subjects. We did not observe any significant difference in our dataset. In the future, we aim to conduct a larger study to explore this further.
% \noindent \textbf{Non-actor Population Considerations}
% Only 40\% of the participants found it easy to express different emotions using the same phrase, highlighting a key challenge in emotion recognition for non-actors. While existing solutions often rely on video capture or show limited accuracy with non-actor datasets, our IMU and audio-based multi-modal approach extracts articulatory motion and audio features to robust emotion recognition.

% \noindent \textbf{Extending to unseen emotions.} We conducted a small feasibility study where one participant expressed 10 phrases in calm (which was not in our dataset). We obtained the valence arousal outputs from our model and placed the phrases on the valence-arousal plot in \ref{fig:pca_plot}. Our results were promising, with

% \subsection{Limitations}
 \noindent \textbf{Limitations.} We identify 3 limitations of \system{} that remain to be addressed. (1) We asked the users to enact emotions during phrase collection. In the future, we will explore using higher fidelity modalities like EEG and explore the correlation between self-reported emotions and brain activity. (2) We did not account for cultural variations among participants, and diversity in facial expressions and jaw motions, requiring user-specific calibration. (3) Although our system needs only 2 minutes of training data per user (Figure~\ref{fig:training_requirement_res}), less than the baseline approaches, this calibration requirement could still present a challenge in its wide-scale adoption.

\section{Conclusion}
~\system{} is the first system to develop a multi-modal model that combines articulatory motion with audio for emotion recognition. By validating our system with a variety of data collected from 20 users (non-actors) across 10 languages, we demonstrate that our system can achieve 93\% emotion recognition. Not only that, we also predict the emotion on the valence-arousal dimensions. We believe that content and language-agnostic emotion recognition can find widespread applications ranging from mental healthcare to HCI.

\bibliography{references}
\bibliographystyle{IEEEtran}

\appendix

\raggedbottom

\section{Dataset Collection}
\label{sec:appendix_dataset}

\subsection{Participants and Study Protocol}

We conducted an IRB-approved study involving 20 participants (12 males and 8 females), who were fluent in English and one or more non-English native languages: Hindi (14), Bengali (4), Marathi (3), Gujarati (2), Mandarin (2), Spanish (1), Tulu (1), Kannada (1), Kutchi (1), and Bhojpuri (1). The study was carried out in a quiet room, free of external noise, in an unrestricted environment where participants were allowed to make unconstrained body movements and hand gestures. Our dataset includes six distinct emotions: Happy, Sad, Fear, Disgust, Anger, and Neutral, aligning with widely used speech emotion datasets in the literature~\cite{livingstone2018ryerson, poria2018meld, jackson2014surrey}.

We asked participants to express specific emotions while reading phrases and responding to prompts. After each recording, participants self-reported their intended emotional state, which we use as labels for training and evaluation. Similar to widely used speech emotion recognition datasets, including IEMOCAP~\cite{busso2008iemocap} and RAVDESS~\cite{livingstone2018ryerson}, as well as recent multimodal emotion recognition systems~\cite{shah2019articulation, wang2022multi}, our data collection protocol relies on prompted emotional expressions. We position our results within the context of controlled emotion elicitation and leave the exploration of more naturalistic elicitation and annotation protocols to future work.

\vspace{-0.2cm}
\subsection{Hardware and Data Collection Setup}

We collected data using a twin-IMU setup, which has been shown to be effective in capturing jaw motion and removing motion artifacts~\cite{srivastava2022muteit}. Figure~\ref{fig:data_collection_comb} (right) shows our data collection prototype. The prototype uses a flexible link that adjusts to any user for wide acceptance, with participants in previous studies~\cite{srivastava2023jawthenticate} reporting it comfortable to wear. The entire system weighs less than 10g, making it lightweight.

The primary IMU was affixed to the temporomandibular joint (TMJ), while the secondary (reference) IMU was secured to the temporal bone using a medical-grade adhesive. We gathered data from a 3-axis accelerometer (100 Hz), a 3-axis gyroscope (100 Hz), and audio (48 kHz), along with the corresponding timestamps. The accelerometer and gyroscope data were streamed to a laptop via Bluetooth, and we used the built-in laptop microphone as the audio source.

We developed a user interface (UI) to record participant keystrokes for start and end timestamps for synchronization across sensor streams. The UI displayed the emotion at the top and the content to be spoken below it, a setup consistent with common practices in SER data collection~\cite{livingstone2018ryerson}. We collected two sessions per user at least three days apart, enabling us to assess the impact of sensor placement variations on performance.

We note that the current implementation is a proof-of-concept system. Our sensing hardware is lightweight and low-power, and the processing can be performed in real time on edge devices such as earables, smart glasses, or AR devices. Future work will focus on fully on-device, real-time implementations and integration into an unobtrusive wearable.

\subsection{Recording Scenarios}

We collected data across scenarios: different linguistic, contextual, and environmental conditions.

\subsubsection{Different Phrases}

Users vocally expressed emotions while speaking 10 phrases (2--3 repetitions in each session) designed to elicit a specific emotional state. This data was used to train the system for our primary emotion recognition task.

\subsubsection{Same Phrase}

Many phrases can be spoken with different emotional tones, but text-only approaches often fail to capture this variability. To assess whether our system is content-agnostic, users were asked to vocally express various emotions while speaking the same set of 10 phrases for each emotion, ensuring that emotional variation was conveyed solely through speech.

\subsubsection{Native Language}

Users were asked to speak two phrases of their choice for each emotion in a non-English native language. This enabled us to evaluate whether the system is content- and language-agnostic.

\subsubsection{Paragraphs}

Users vocally expressed emotions while reading aloud three paragraphs for each of the five target emotions (Happy, Sad, Anger, Fear, and Disgust). This data was used to investigate how longer spoken sequences or snippets from familiar stories impact system performance.

\subsubsection{Conversational Data}

We conducted conversational prompts with five participants designed to elicit everyday emotional expressions. These prompts were designed to encourage participants to express target emotions in a conversational setting while still relying on self-reported labels.

The conversations covered three topics: travel experiences, food preferences, and TV shows. Each category contained questions designed to evoke specific emotions. For example, `What is your favorite TV show?'' was used to elicit happiness, while `What TV show timeline or plot hole annoys you the most?'' was used to evoke anger or frustration.

Participants self-reported their emotional state after each question. When participants reported experiencing multiple emotions simultaneously (e.g., both sadness and anger), we considered the model prediction accurate if it matched any of the self-reported emotions. On average, we collected six samples per emotion.

\subsubsection{Motion and Acoustic Noise Data}

We collected data from six participants in the presence of motion noise, wherein participants were free to walk around the room during data collection. Each participant expressed each emotion 20 times.

In addition, we collected data with five users in the presence of acoustic noise, playing ambient restaurant sounds, rock music, and hallway conversations at approximately 80~dB. Each participant recorded five samples for each emotion under these noise conditions. This approach helped us evaluate our system's performance under typical movement and acoustic noise conditions that users would experience in daily life.

\subsection{Label Validation}

To assess label consistency, three independent graduate student annotators reviewed a subset of the data consisting of 20 samples per emotion, spanning 25 unique phrases and five same-phrase samples. We observed a Pearson correlation coefficient of 0.93 between annotator labels and participant self-reports, suggesting strong agreement despite the subtle nature of non-actor emotional expressions.

%\clearpage
%\input{appendix}
\balance
\end{document}